# Gate-tunablePhotoresponse Time in BlackPhosphorus-MoS$_2$Heterojunctions

*Thayer S. Walmsley[1], Bhim Chamlagain[2], Tianjiao Wang[3], Zhixian Zhou[2], and Ya-Qiong Xu[1,3]*

[1]Department of Physics and Astronomy, Vanderbilt University, Nashville, TN 37235, USA

[2]Department of Physics and Astronomy, Wayne State University, Detroit, MI 48201, USA

[3]Department of Electrical Engineering and Computer Science, Vanderbilt University, Nashville, TN 37235, USA

**Abstract**

We study the rise and decay times in BP-MoS$_2$ heterojunctions through gate- and wavelength-dependent scanning photocurrent measurements. Our results have shown that the Schottky barrier at the MoS$_2$-metal interface plays an important role in the photoresponse dynamics of the heterojunction. When the MoS$_2$ channel is in the on-state, photo-excited carriers can tunnel through the narrow depletion region at the MoS$_2$-metal interface, leading to a short carrier transit time. A response time constant of 13 μs has been achieved in both the rising and decaying regions regardless of the incident laser wavelength, which is comparable or higher than those of other BP-MoS$_2$ heterojunctions as well as BP and MoS$_2$ based phototransistors. On the other hand, when the MoS$_2$ channel is in the off-state the resulting sizeable Schottky barrier and depletion width make it difficult for photo-excited carriers to overcome the barrier. This significantly delays the carrier transit time and thus the photoresponse speed, leading to a wavelength-dependent response time since the photo-excited carriers induced by short wavelength photons have a higher probability to overcome the Schottky barrier at the MoS$_2$-metal interface than long wavelength photons. These studies not only shed light on the fundamental understanding of photoresponse dynamics in BP-MoS$_2$ heterojunctions, but also open new avenues for engineering the interfaces between two-dimensional (2D) materials and metal contacts to reduce the response time of 2D materials based optoelectronic devices.

**Introduction**

The successful isolation and characterization of graphene opened up a new research field into two-dimensional (2D) materials, thus prompting the exploration and exploitation of their unique mechanical, electrical, and optical properties.[1-2] However, the gapless nature of graphene limits its application in photovoltaics and photodetection. This has fueled the search for other 2D materials that have appreciable band gaps. A class of materials known as transition metal dichalcogenides (TMDs) have attracted attention due to their remarkable electrical transport properties and strong light-matter interaction, with molybdenum disulfide ($MoS_2$) being one of the most notable in that class of materials.[3-6] A single layer of $MoS_2$ has a direct band gap of 1.9 eV, while its bulk material has an indirect band gap around 1.3 eV.[7] Despite $MoS_2$ phototransistors demonstrating high photoresponsivities greater than 100 mA/W under 633 nm illumination, their sizeable band gap significantlylimits their potential applications in near-infrared (NIR) spectral region.[3, 8-9] Recently, another 2D semiconducting material, black phosphorous (BP), has exhibited a direct band gap ranging from ~ 0.3 eV in bulk to 2 eV for a monolayer structure, thus allowing few-layer BP to be used for NIRphotodetection.[10-11] Additionally, 2D materials offer the advantage of creating vertical heterostructures by simply stacking these materials together without the traditional concerns of lattice mismatch.[12] This allows for novel semiconductor engineering and optimization of device properties.[13-14] For instance,a vertical heterojunction formed by stacking BP and $MoS_2$ flakes together displays a high photoresponsivity up to 22.3 A/W under 532 nm illumination and 153.4 mA/W under 1550 nm illumination,[14-15] while the photoresponsivity ofBP based devices is around 4.8 mA/W under 640 nm illumination.[11]

Aside from photoresponsivity, the photoresponse time is another key figure of merit for photodetectors. A great variety of response times have been reported for these 2D materials and under various conditions, ranging from 40 μs to a few milliseconds for BP[11, 16-17] and from 70 μs to over a second for $MoS_2$.[3, 8-9, 18] A recent report has shown that the rise time for BP-$MoS_2$ heterojunctions is ~15 μs, while its decay time is around 70 μs.[15] It is well-known that the response time of photodetectors is limited by the carrier transit time,[19] which depends on electrical transport properties of individual channels and junctions. Though some workshave investigated engineering channel length to control the rise time for BP and $MoS_2$phototransistors,[17, 20] little attention has been paid to exploring how gate voltages impact carrier transit time in either the BP or $MoS_2$channel as well as the role that the incident laser wavelength has on the rise and decay timesofBP-$MoS_2$ heterojunctions. Understanding the mechanisms and the conditions that give rise to certain response time behavior is vital to the broad applicability of 2D vertical heterojunctions use as photodetectors.

Here we systematically investigate photoresponse dynamics in BP-$MoS_2$ heterojunctions through gate- and wavelength-dependent scanning photocurrent measurements. We have found that the Schottky barrier at the $MoS_2$-metal interface plays a significant role in the rise and decay times of the heterojunction. When the $MoS_2$ channel is in the on-state, photo-excited carriers can tunnel through the narrow depletion region at the $MoS_2$-metal interface, leading to a short carrier transit time. For both the rising and decaying region a response time constant of 13 μs has been achieved, irrespective ofthe incident laser wavelength.Our reported response time constants are comparable or better than those of other BP-$MoS_2$ heterojunctions as well as BP and $MoS_2$ based phototransistors. In contrast, when the $MoS_2$ channel is in the off-state, the sizeable Schottky barrier and depletion width make it difficult for photo-excited carriers to overcome the barrier, which significantly delays the carrier transit time and thus the photoresponse speed. Under this condition since photo-excited carriers induced by short wavelength photons possess higher energy, they have a higher probability to overcome the Schottky barrier at the $MoS_2$-metal interface resulting in a relatively fast response in comparison with that of carriers excited by the incident light with a long wavelength. Consequently, the rise and decay time constants of the heterojunction increase with increasing incident laser wavelength. These fundamental studies not only provide an in-depth understanding of the photoresponse dynamics in $BP$-$MoS_2$ heterojunctions, but also offer a new way to reduce the response time of 2D heterojunctions by engineering 2D material-metal interfaces.

**Results and Discussion**

Figure 1a illustrates the schematic diagram of a $BP$-$MoS_2$ heterojunction. In our experiment, $MoS_2$ thin flakes were first mechanically exfoliated from its bulk crystal onto a degenerately-doped 290 nm $SiO_2$/Si substrate. Next, a mechanically-exfoliated BP flake on a polydimethylsiloxane stamp was selected to be placed atop a selected $MoS_2$ flake on the $SiO_2$/Si substrate to form a $BP$-$MoS_2$ heterojunction. Metal electrodes were then patterned onto the device using electron beam lithography followed by the deposition of 5 nm of Ti and 40 nm of Au. Figure 1b shows an optical image of a typical $BP$-$MoS_2$ heterojunction, where BP and $MoS_2$ flakes are outlined by blue and purple dashed lines, respectively, and the electrodes are outlined in grey dashed lines. A Park-Systems XE-70 noncontact atomic force microscopy was utilized to determine the thickness of the BP (~

14 nm) and $MoS_2$ (~ 3 nm) flakes. We measure the electrical transport properties of the junction under high vacuum (~$10^{-6}$ torr) at room temperature in a Janis ST-500 microscopy cryostat, where gate voltages were applied to the Si substrate to adjust the carrier concentration in each material. Figure 1c presents the gate-dependent transport characteristics of two individual flakes, in which the BP and $MoS_2$ flakes show typical p-type and n-type behavior at a zero gate voltage, respectively. The *I-V* characteristics of the BP–$MoS_2$ junction were also explored at various gate voltages as seen in Figure 1d. Consistent with gate-tunable transport properties, strong rectification of the drain current can be seen at a zero gate voltage where unintentional doping occurs. As the gate voltage is modified, a decrease in these rectification properties is observed due to the modulation of carrier concentrations. These rectification properties are similar to traditional p-njunctions except the electrical tunability in vertical van Der Waals heterojunctions between two atomically-thin materials is likely attributed to tunneling-assisted interlayer recombination given the absence of a vertical depletion region.[13]

Aside from the electrical transport measurements, the optoelectronic properties of the device were investigated. A spatially-resolved photocurrent map was generated using scanning photocurrent microscopy. This was done using a continuous wave laser beam (NKT Photonics SuperK Supercontinuum Laser) that was expanded and then focused by a 40X Olympus objective (N.A. = 0.6) into a diffraction-limited spot (~1 μm) and scanned over the device using piezo-controlled mirrors with nanometer-scale spatial resolution. Under 650 nm illumination, the reflection and photocurrent images of a BP-$MoS_2$ heterojunction were collected simultaneously to locate the position of photocurrent signals. As shown in Figure 2, a strong photocurrent response ($I_{pc} = I_{laser} - I_{dark}$) was observed in the overlap region between BP and MoS$_2$ flakes, where the built-in electric field can efficiently separate the photo-excited electron-hole pairs (EHPs) to generate photocurrent signals.

We further explore the photoresponse dynamics of BP-MoS$_2$ heterojunctions with various incident light wavelengths and gate voltages through temporally resolved measurements. To do so, an optical chopper was utilized to mechanically apply on/off light modulation to the laser beam while photocurrent signals were collected as a function of time. As shown in Figures 3a, the photoresponse time of the heterojunction under 650 nm illuminations (black) at a gate voltage of -20 V is faster than that excited by 1000 nm photons (red). To obtain rise or decay time constants ($\tau_r$ or $\tau_d$) for the device, a single exponential function was applied to fit the rising or decaying region of the curve, respectively. (Figure 3b). The rise time constant is defined as $\tau_r = t_r/\ln(9)$, where $t_r$ is the time that it takes the device to rise from 10% to 90% of the maximum signal. As shown in Figure 3c, the rise time constant of the junction gradually increases from 35 µs to 90 µs when the illumination wavelength increases from 650 nm to 1000 nm and a -20 V gate voltage is applied, while the rise time constant of the device remains constant with increasing wavelength for gate voltages of both 0V (~ 22 µs) and 20V (~13 µs). Moreover, the rise time constant increases with decreasing gate voltage regardless of the wavelength.

The observed behavior of the heterojunction can be explained by energy band diagrams in Figure 4. When the MoS$_2$ channel is in the off-state (-20 V), there exists a sizeable Schottky barrier and depletion region located at the MoS$_2$-metal interface, which make it difficult for carriers to overcome the barrier and to be collected by the metal electrode, leading to a long carrier transit time. When the Fermi level of MoS$_2$ moves close to its conduction band, the band offset ($\phi_0 = E_c - E_F$) changes linearly with the back gate voltage $V_G$: $\delta\phi_0 = e\alpha V_G$, where $E_c$ is the minimum of the MoS$_2$ conduction band, $E_F$ is its Fermi energy, $e$ is the electron charge, and $\alpha$ is a numerical constant that measures how effectively the gate voltage modulates the band energy.[21-22]The shut-off gate voltage observed in the transfer characteristic of the MoS$_2$ channel is about -10 V, indicating that the Fermi level is close to the minimum of the MoS$_2$ conduction band under this gate voltage.[23-24]On the other hand, when the Fermi level is very close to the conduction band of MoS$_2$(0 V or 20 V), the depletion width will become very narrow. As a result, the carriers can easily flow from MoS$_2$ to the metal electrode due to tunneling effect through a narrow depletion region, leading to an ohmic-like contact.[25]

Under 1000 nm illumination, the photocurrent signals are mainly attributed to the direct band gap transition in BP since the incident photons cannot provide sufficient energy to excite electrons in the MoS$_2$ from its valence band to its conduction band(Figure 4a). Based on the polarity of photocurrent responses, we can identify that photo-excited electrons in BP flow from BP to MoS$_2$ due to the conduction band offset at the BP-MoS$_2$ interface.[26]These photo-excited electrons will then pass through the MoS$_2$ channel to the metal electrode. When the MoS$_2$ channel changes from the on-state (20 V)to the off-state (-20 V) the depletion width at the MoS$_2$–metal interface increases, making it more difficult for photo-excited electrons to overcome the interface barrier and thus leading to a long carrier transit time. Therefore, the photoresponse time increases when the MoS$_2$ channel is in the off-state since the carrier transit time is the key factor that limits the photoresponse time.[19]Similar results have been reported in MoS$_2$ phototransistors, where a large decrease in the photoresponse time was observed when the MoS$_2$channel was turned on.[8, 27]Moreover, photons with a shorter wavelength and thus higher energy can transfer more energy to EHPs, leading to high energy photo-excited electrons with a higher probability to overcome the Schottky barrier at the $MoS_2$-metal interface when the $MoS_2$ channel is in the off-state.[3] The subsequent flow of higher energy photo-excited electrons to the external circuit to generate the photocurrent has a shorter transit time and thus faster photoresponse time. However, when the $MoS_2$ channel is turned on photo-excited electrons can tunnel through the narrow depletion region; thus the photoresponse time does not depend on the energy of these photo-excited electrons (or the wavelength of the incident light).

Under 650nm illumination, the electrons in both BP and $MoS_2$ are able to be excited from their valence to conduction bands. Due to the Fermi level alignment of BP and $MoS_2$, the valence band maximum of BP is much higher than that of $MoS_2$ (Figure 4b); therefore, photo-excited holes in the valence band of $MoS_2$ can inject into the BP and then pass through the BP channel to the electrode resulting in strong photocurrent signals with the same polarity as those induced by the photo-excited electrons in BP. For photo-excited holes, the barrier at the BP-metal interface is negligible due to a high work function of Au (Figure 4b), having a limited effect on the carrier transit time. Thus, there is no obvious photoresponse time delay observed when the BP channel is in the off-state ($V_G > 1\ V$).

We also investigate the decay process of the BP-$MoS_2$ heterojunction after turning off the incident light. As shown in Figure 4d, the decay time follows a similar gate- and wavelength-dependence as the rise time, but is slightly longer than the corresponding rise time when the $MoS_2$ channel is turned off. Here, the longer decay time is likely related to the defect/impurity trapping present at the $MoS_2$-metal interface.[28] When the light is on, photo-excited carriers get trapped in these states. Once the light turns off, the photoresponse cannot fully vanish until the trapped carriers are released leading to a longer decay time. However, when the $MoS_2$ channel is in the on-state, the photo-excited electrons can tunnel through the narrow depletion region without filling the trapping at the interface. Thus, there is no delay for the decay time under this circumstance. In our experiment, a decay time of 13µs regardless of the incident laser wavelength has been achieved with a gate voltage of 20 V, which is several timesfaster than that ofother BP-$MoS_2$ heterjunctions[15]as well as BP and $MoS_2$ based phototransistors.[3, 18, 27, 29] Here, the rise and decay times are mainly limited by the sizable Schottky barriers at the $MoS_2$-metal interface. Carefully engineering the $MoS_2$-metal contact by adding heavily-doped TMDs to generate extremely narrow depletion regions[25, 30] or using gated-graphene contact to achieve negligible Schottky barriers[31] may further improve the photoresponse time of BP-$MoS_2$ heterojunctions.

**Conclusion**

In conclusion, spatially-resolved photocurrent microscopy has been utilized to study thephotoresponse time of BP-$MoS_2$ heterojunctions excited by incident light of wavelengths ranging from 650 nm to 1000 nm. We have found that the Schottky barrier at the $MoS_2$-metal interface plays an important role in the rise and decay time of the heterojunction. A longer carrier transit time and thus a slow photoresponse speed is observed when the $MoS_2$ channel is turned off due tothe sizeable Schottky barrier and depletion width that make it difficult for photo-excited carriers to overcome the barrier at the $MoS_2$-metal interface.Moreover, our results have shown that the rise and decay time of the heterojunction increase with increasing incident laser wavelength under this condition. Here, photo-excited carriers induced by short wavelength photons possess higher energy, resulting in a higher probability to overcome the Schottky barrier at the $MoS_2$-metal and thus a relatively fast response time. While at longer wavelengths the lower energy incident photons are unable to overcome the barrier and thus have an even longer response time. On the other hand, when the $MoS_2$ channel is in the on-state, photo-excited carriers can tunnel through the narrow depletion region at the $MoS_2$-metal interface, leading to a short carrier transit time independent of the wavelength of the incident photon. A response time constant of 13 μs has been achieved in both rising and decaying regions regardless of the incident laser wavelength, which is comparable or higher than those of other BP-$MoS_2$ heterojunctions as well as BP and $MoS_2$ based phototransistors. Our work expands the understanding of photoresponse dynamics in 2D material heterojunctions and offernew insight into the engineering of the interfaces between 2D materials and metal electrodes to reduce their photoresponse times, opening a door for future high-speed electronic and optoelectronic applications.

**Acknowledgement**

This work was supported by the National Science Foundation (ECCS-1055852, ECCS-1810088, and DMR-1308436).

**Additional Information**

The authors declare no competing financial interests.

**Figures**

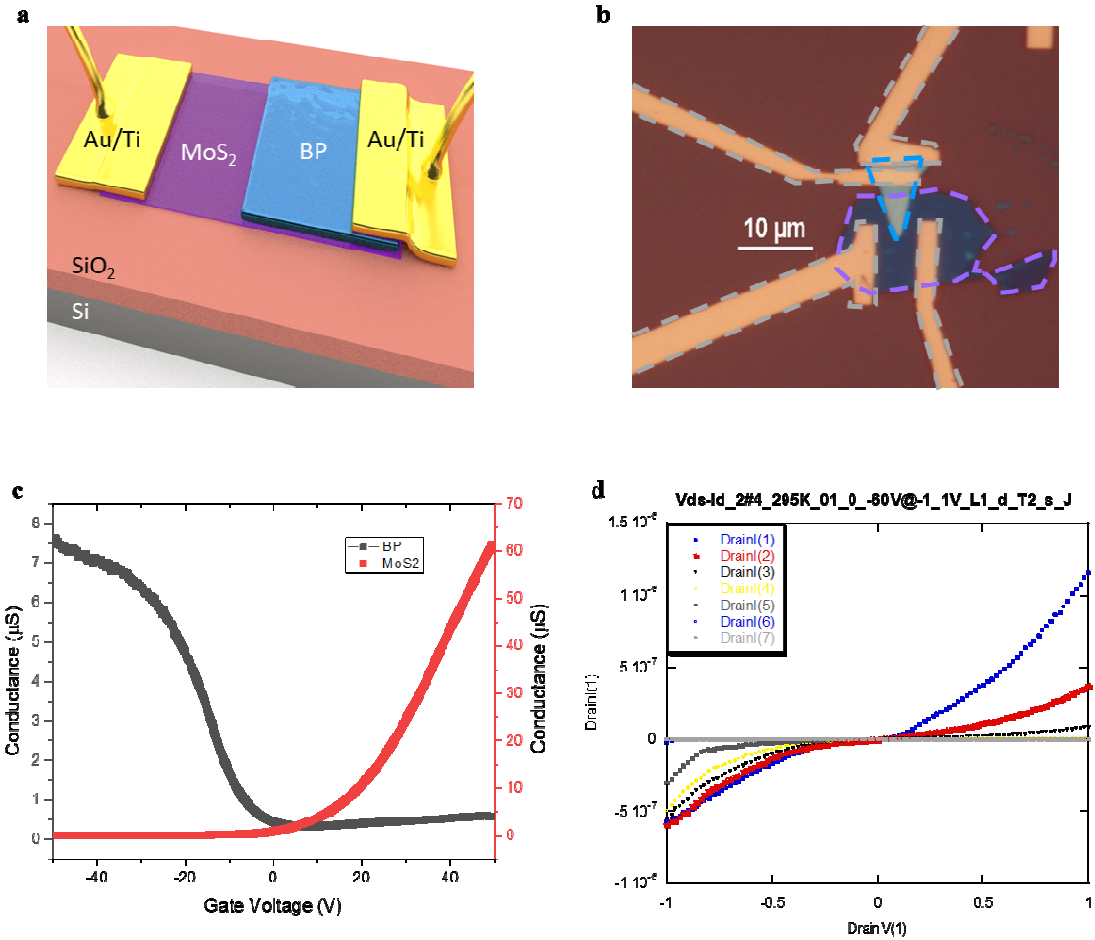

**Figure 1.** (a) A schematic diagram and (b) an optical image of a typical BP-MoS$_2$ heterojunction. The grey dashed lines outline the Au electrodes. The blue and purple dashed lines outline the BP and MoS$_2$ flake, respectively. (c) Gate-dependent transport characteristics of the BP (black) and MoS$_2$ (red) channelsof the device, respectively. (d) The electrical transport properties of the heterojunction are shown.

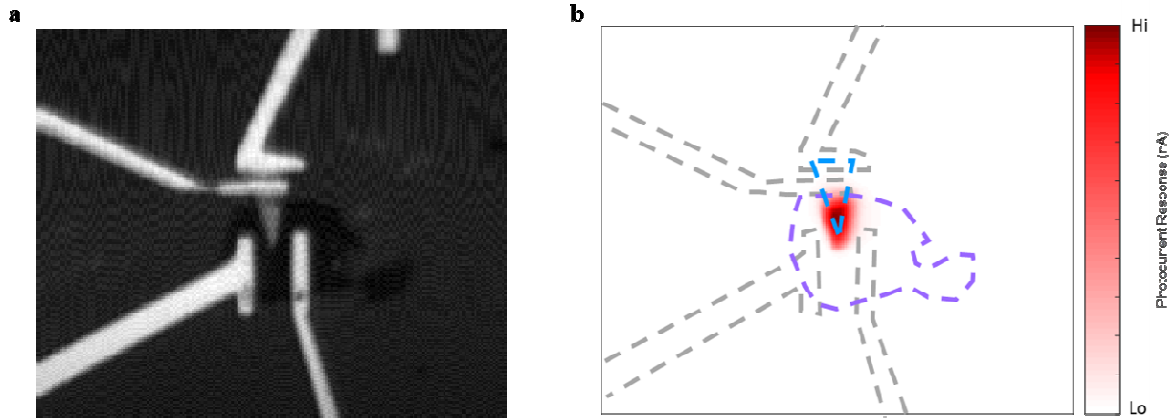

**Figure 2.** (a)The reflection and (b) scanning photocurrent images of the BP-MoS$_2$ heterojunction under a zero gate voltage. The grey dashed lines show the outline of the electrodes. The blue and purple dashed lines represent BP and MoS$_2$ flakes, respectively.

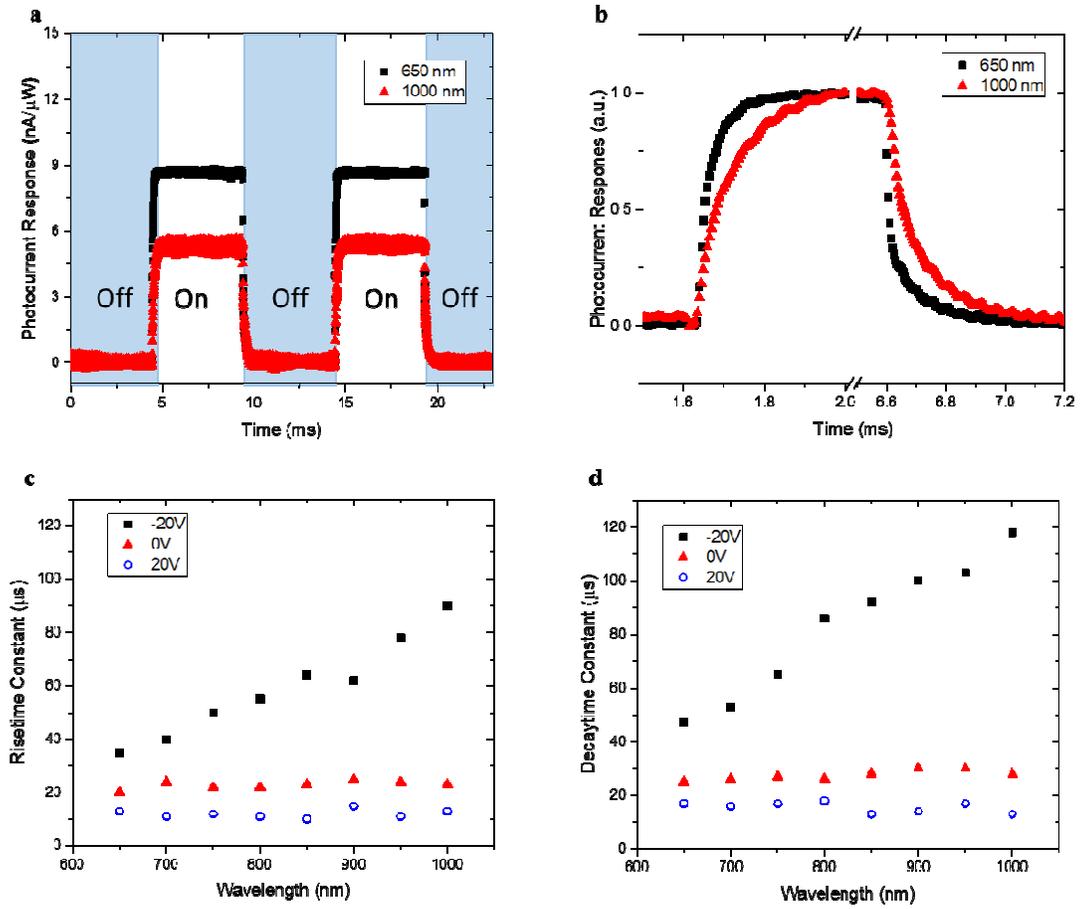

**Figure 3.** (a) The photocurrent temporal response of the BP-MoS$_2$ heterojunction under a gate voltage of -20V for laser illumination wavelengths of 650nm (black) and 1000nm (red), respectively. (b) The normalized photocurrent response of (a) to elucidate the rising and decaying curves. (c) Rise and (d) decay time constants as a function of the wavelength of the incident light under different gate voltages: -20V (black square), 0V (red triangle), and 20V (blue open circle).

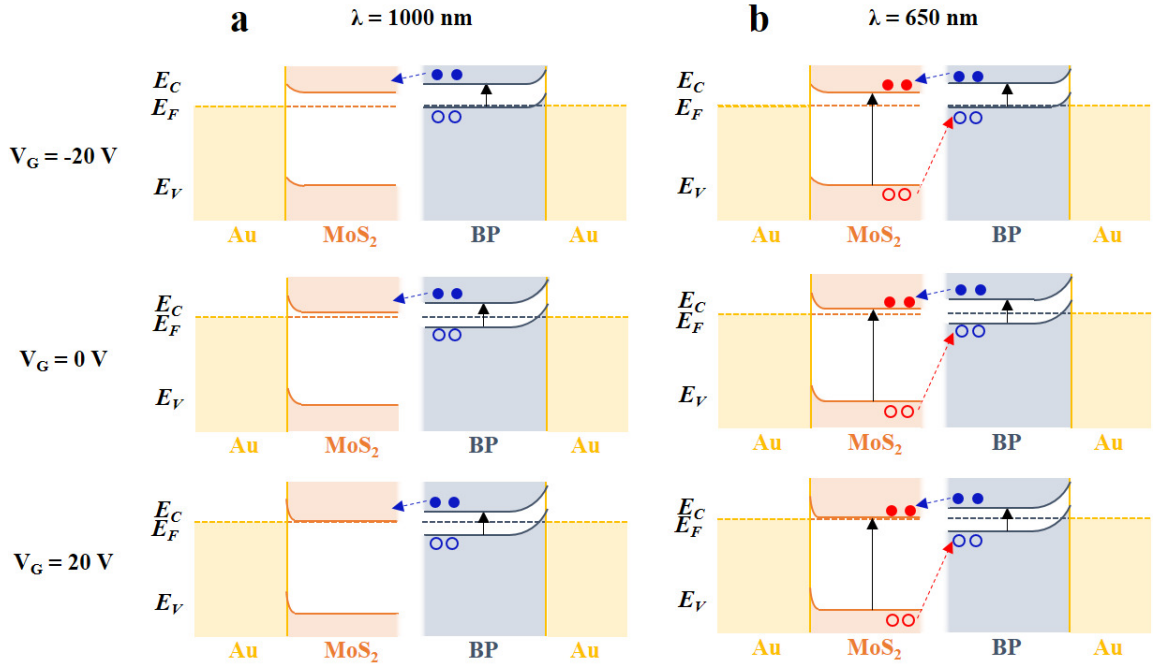

**Figure 4.** The band diagrams for BP-MoS$_2$ heterojunctions under (a) 1000 nm and (b) 650 nm illumination with different gate voltages.